\documentclass[acmlarge]{acmart}
\pdfoutput=1 
\AtBeginDocument{%
	\providecommand\BibTeX{{%
			\normalfont B\kern-0.5em{\scshape i\kern-0.25em b}\kern-0.8em\TeX}}}

\setcopyright{acmcopyright}
\copyrightyear{2021}
\acmYear{2021}
\acmDOI{10.1111/11.11}

\acmJournal{HEALTH}
\acmVolume{11}
\acmNumber{1}
\acmArticle{1}
\acmMonth{1}

\begin{document}
	
	\title{Transfer Learning for Human Activity Recognition using Representational Analysis of Neural Networks}
	
	\author{Sizhe An}
	\affiliation{%
	  \institution{University of Wisconsin-Madison}
	  \streetaddress{1415 Engineering Dr}
	  \city{Madison}
	  \state{Wisconsin}
	  \country{USA}
	  \postcode{53706}}
	\email{sizhe.an@wisc.edu}
	
	\author{Ganapati Bhat}
	\affiliation{%
	  \institution{Washington State University}
	  \streetaddress{355 NE Spokane St}
	  \city{Pullman}
	  \state{Washington}
	  \country{USA}
	  \postcode{99164}}
	\email{ganapati.bhat@wsu.edu}
	
	\author{Suat Gumussoy}
	\affiliation{%
	  \institution{Siemens Corporate Technology}
	  \country{USA}}
	\email{suat.gumussoy@siemens.com}
	
	\author{Umit Ogras}
	\affiliation{%
	  \institution{University of Wisconsin-Madison}
	  \streetaddress{1415 Engineering Dr}
	  \city{Madison}
	  \state{Wisconsin}
	  \country{USA}
	  \postcode{53706}}
	\email{uogras@wisc.edu}	
	
	
	\begin{abstract}
\label{sec:abstract}
Human activity recognition (HAR) has increased in recent years due to its applications in mobile health monitoring, activity recognition, and patient rehabilitation. The typical approach is training a HAR classifier offline with known users and then using the same classifier for new users.
However, the accuracy for new users can be low with this approach if their activity patterns are different than those in the training data. 
At the same time, training from scratch for new users is not feasible for mobile applications due to the high computational cost and training time.
To address this issue, we propose a HAR transfer learning framework with two components. First, a representational analysis reveals common features that can transfer across users and user-specific features that need to be customized.
Using this insight, we transfer the reusable portion of the offline classifier to new users and fine-tune only the rest. Our experiments with five datasets show up to 43\% accuracy improvement and 66\% training time reduction when compared to the baseline without using transfer learning. Furthermore, measurements on the hardware platform reveal that the power and energy consumption decrease by 43\%  and 68\%, respectively, while achieving the same or higher accuracy as training from scratch.
\end{abstract}
	%
	%
	%
	%
	\maketitle
	
	\section{Introduction}

Human activity recognition is a critical component in a range of health and activity monitoring applications~\cite{maetzler2016clinical,heldman2017telehealth,bort2014measuring}. 
It provides valuable insight into movement disorders by allowing health professionals to monitor their patients in a free-living environment~\cite{daneault2018could,espay2016technology,OpenHealth}. 
HAR is also the first step towards understanding gait parameters, such as step length and gait velocity, which are also used in movement disorder analysis and rehabilitation~\cite{schlachetzki2017wearable,yang2019review}.
In addition, HAR is used for obesity management and promoting physical activity among the public. Due to these high-impact applications of HAR, it has received increased research attention in recent years~\cite{shoaib2015survey,wang2019deep}.

\begin{figure*}[t]
    \centering
    \includegraphics[width=1\linewidth]{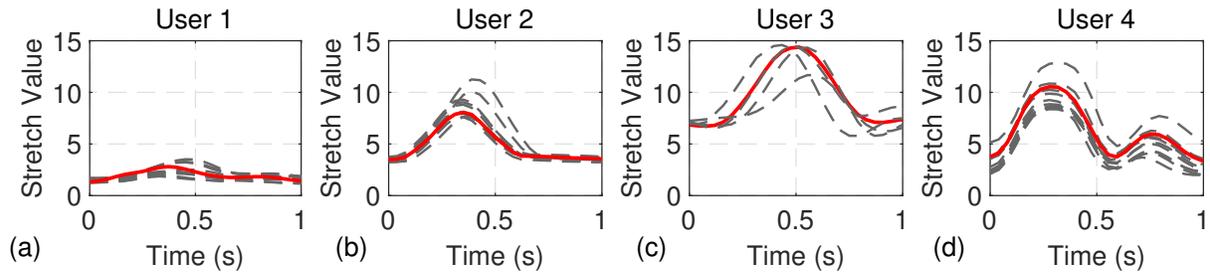}
    \caption{Comparison of stretch sensor~\cite{o2014stretch} data of four users for a single step during walk. There is a significant change in both the range of values and data pattern. The grey dashed lines show different instances of the same activity, while the red line shows a \textit{representative} activity window for each user.}
    \label{fig:motivation}
    \vspace{-5mm}
\end{figure*}

Most HAR techniques start with collecting sensor data from users available at design time~\cite{lara2013survey}. This data is used to train a classifier for the activities of interest.
Then, the trained classifier is used by new users, whose data is not available for  training.
This approach assumes that the HAR classifiers can be transferred across different user sets. 
However, this assumption may not hold in general as activity patterns can change with age, gender, and physical condition. Hence, lack of model personalization can limit the accuracy~\cite{lin2020model}. For instance, Figure~\ref{fig:motivation} shows the stretch sensor data during walking for four users in our experimental dataset.
There is a significant variation both in the range of sensor values and the data patterns. 
Furthermore, the activity patterns may vary over time even for a given user due to progression of symptoms, injury, or other physical changes.
These variations can significantly reduce the recognition accuracy for new users as we demonstrate in this paper.
Therefore, \textit{classifiers designed offline must adapt to changing data patterns of new and existing users to achieve high classification accuracy}. 

Training classifiers from scratch for new users is expensive due to data storage and computational requirements. It is challenging especially for low-power wearable and mobile devices, which are the most common platforms for HAR~\cite{shoaib2015survey,lara2013survey}. 
Moreover, training from scratch for a particular user loses generalization capability and robustness due to overfitting~\cite{morcos2018insights}.
In contrast, transfer learning with a common feature set can carry over generalization capabilities from offline stage and fine-tune user-specific features after deployment to improve training efficiency.


This paper first demonstrates that HAR classifiers designed offline cannot be transferred as a whole to an arbitrary set of users. 
Then, it presents the first systematic study to determine how to transfer the offline knowledge and adapt HAR classifiers to individual users. We show not only the feasibility of transferring the offline classifier, but also determine which parts can be transferred to minimize the \textit{time and energy} required for customization. Furthermore, we demonstrate these benefits on a mobile hardware board using five datasets.
This study is performed using convolutional neural networks~(CNNs) due to their ability to produce broadly applicable features from raw input data. 
We use canonical correlation analysis (CCA)~\cite{morcos2018insights} to evaluate the distance between the layers of CNNs trained with different sets of users. 
Our analysis leads to a theoretically grounded and practical HAR classifier framework validated on hardware.
It achieves high accuracy and significant savings in training time 
by fine-tuning the deeper layers
that differentiate users while transferring the earlier layers. 

The proposed approach is validated using five public datasets~\cite{w_har,anguita2013public,reyes2016transition,kwapisz2011activity,micucci2017unimib}. 
We first divide the users in each dataset into multiple clusters to evaluate the effect of transfer learning for unseen user clusters. 
In a challenging scenario, the clusters have to be as separated as possible such that the classifier can only learn the patterns from the users in the training set.
For instance, the four users in Figure~\ref{fig:motivation} belong to different clusters. To this end, we generate user clusters both randomly (for average-case) and using k-means clustering~\cite{friedman2001elements} (for the most challenging scenario).
Next, a CNN classifier is trained for each user cluster. After analyzing the similarity between the CNNs trained with different user clusters, the weights are transferred between user clusters and the dissimilar layers are fine-tuned. 
Extensive evaluations for the w-HAR dataset show that in the worst case, transferring weights and fine-tuning the last layer achieves accuracy that is same as the accuracy obtained by training from scratch while reducing the computation during training by 66\%. 
Finally, we implement the proposed approach on the Nvidia Jetson Xavier-NX board~\cite{Xavier}. Our experiments show up to 68\% energy and 43\% power consumption savings. These results demonstrate the practicality of the proposed technique on low-power edge devices.

In summary, the major contributions of this paper are:
\vspace{-1mm}
\begin{itemize}\itemsep-1pt
    \item An empirical demonstration which shows that HAR classifiers designed offline cannot be transferred to an arbitrary set of users,
    \item A systematic analytical study that reveals that deeper (typically last two) layers of HAR classifiers capture user-specific information, while the first three to four layers provide general features,
    \item Extensive experimental evaluations using five datasets that show up to 43\% and on average 14\% higher accuracy compared to the accuracy without using transfer learning for new users.
    
    \item Hardware experimental evaluations that demonstrate up to 68\% energy and 43\% power consumption savings.
\end{itemize}

	\section{Related research}
\textbf{Transfer learning:} Transfer learning aims to leverage the information learned in one domain to improve the accuracy in a new domain~\cite{pan2010survey}.
The information used for transfer includes the weights of a classifier~\cite{schwaighofer2005learning,evgeniou2004regularized,feng2019few}, features~\cite{raina2007self}, and instances of data~\cite{Dai:2007:BTL:1273496.1273521}.
The transfer can occur between different applications or between different scenarios of a single application~\cite{pan2010survey}. 
A popular example of transfer learning between applications is medical imaging~\cite{shin2016deep,salem2018ecg} where CNNs trained for classifying ImageNet~\cite{deng2009imagenet} are adapted to classify medical images.
Similarly, transfer learning for new scenarios includes adaptation to a new device~\cite{akbari2019transferring}, classes~\cite{blanke2010remember}, or users~\cite{hachiya2012importance}. 


One of the fundamentals aspects of transfer learning is identifying what information to transfer. Prior work addressed parameter transfer~\cite{oquab2014learning}, feature representation transfer~\cite{argyriou2007multi,blitzer2006domain}, and data instance transfer~\cite{Dai:2007:BTL:1273496.1273521}. We focus on parameter transfer literature since the proposed approach uses parameter transfer as well.
~\citet{oquab2014learning} design a method to reuse layers trained with one dataset to compute mid-level image representation for images in another dataset. ~\citet{yosinski2014transferable} empirically quantify the generality versus specificity of neurons in each layer of a deep convolutional neural network for the ImageNet~\cite{deng2009imagenet} dataset. They show that the features in the initial layers are general in that they are applicable to multiple image recognition tasks.
~\citet{morcos2018insights} directly analyze the hidden representations of each layer in CNN by using CCA, which enables comparing learned distributed representations between different NN layers and architectures. 

Transfer learning has been applied successfully in fields such as medical imaging classification
and computer vision. 
~\citet{salem2018ecg} present an approach to transfer a CNN from image classification to electrocardiogram (ECG) signal classification domain.
Similarly,~\citet{raghu2019transfusion} explore properties of transfer learning from natural image classification networks to medical image classification.
~\citet{quattoni2008transfer} show that prior knowledge from unlabeled data is useful in learning a new visual category from few examples. 
The authors develop a visual-category learning algorithm called sparse prototype learning that can learn an efficient representation from a set of related tasks while taking advantage of unlabeled data.

\textbf{Transfer learning for HAR:} 
Research on HAR has increased in recent years due to its potential in applications such as movement disorders, obesity management, and remote patient monitoring~\cite{maetzler2016clinical,heldman2017telehealth,bort2014measuring}.
One of the challenges in HAR algorithms is that the data available at design time may not be representative of the activity patterns of new users. As a result, the accuracy can degrade for new users~\cite{ding2019empirical}. Recent research has used transfer learning to address this issue~\cite{rokni2018personalized,cook2013transfer,ding2019empirical}.
The survey by ~\citet{cook2013transfer} presents how different types of transfer learning have been used for HAR. 
~\citet{ding2019empirical} perform an empirical study to analyze the performance of transfer learning methods for HAR and find that maximum mean discrepancy method is most suitable for HAR.
A CNN-based method to transfer learned knowledge to new users and sensor placements is presented in~\cite{chikhaoui2018cnn}. The authors empirically determine the number of layers to transfer based on the accuracy obtained after transferring.
However, this method is not scalable since the training has to be repeated each configuration of the transfer.
~\citet{rokni2018personalized} use transfer learning to personalize a CNN classifier to each user by retraining the classification layer with new users.
However, the authors do not provide any insight into the number of layers that can be transferred between users and how it benefits the learning for new users.
In contrast, we perform representational analysis using CCA to determine layers that need to be fine-tuned for new users.
   
Our work proposes a complete transfer learning framework for HAR using representational analysis of CNNs.
We start with clustering the users such that they are as separated as possible. This clustering allows us to test the \textit{robustness} of our approach.
Then, we analyze the hidden representation of different layers of a CNN by using CCA similarity.
Using insights from the CCA similarity analysis, we fine-tune specific layers of the network to adapt to new users. The proposed approach is evaluated on five datasets with both manually and randomly generated clusters. 


\section{Human activity recognition framework}
\label{sec:method}
Human activity recognition aims to identify physical activities, such as walking, sitting, and standing. 
Most HAR approaches start with data from wearable inertial sensors or a smartphone to record the data when the user is performing the activities of interest.\footnote{Video cameras are also used for HAR, but we focus on HAR using wearable sensors and smartphones.}
After collecting the sensor data, the next step is to pre-process and segment the sensor data for feature generation. The most common approach in literature for segmentation is to divide the data into one to ten-second windows with a 50\% overlap between consecutive windows~\cite{anguita2013public,wang2016comparative,kwapisz2011activity}.
Then, the data within each window is processed to generate the features for the classifier. A variety of classifiers, such as neural networks, decision trees~(DT), random forest~(RF), support vector machine~(SVM), and k-nearest neighbors~(KNN), have been used for HAR. 
Since previous approaches use different datasets and activities for evaluation, a direct comparison to our approach is not possible. For a fair comparison, we use the five datasets in this paper on commonly used HAR algorithms.
Table~\ref{tab:accuracy} shows the accuracy of these approaches on the w-HAR dataset. The first row shows the accuracy when the classifier is tested on the users available during training and we refer it to the classifier accuracy with users in test set.  The second row shows the accuracy for new users and we refer it to the classifier accuracy with users in validation set.
All the classifiers achieve a high accuracy on the users available for training, however the accuracy drops significantly when tested on new users. The accuracy drop is minimum for the CNN classifier. This shows that CNN has more potential to generalize by producing broadly applicable features in the convolutional layers. Furthermore, CNNs can be easily fine-tuned at runtime for new subjects using the proposed transfer learning approach. This is much more challenging for other approaches. Therefore, we use CNNs in this paper.


\begin{table}[h]
\centering
\caption{Comparison of classifiers}
\label{tab:accuracy}
\begin{tabular}{@{}llllll@{}}
\toprule
 & KNN & SVM & DT & RF & CNN \\ \midrule
Classifier accuracy with users in the test set (\%)       & 90  & 89  & 97 & 96 & 98  \\
Classifier accuracy with users in the validation set (\%) & 58  & 61  & 52 & 42 & 76  \\ \bottomrule
\end{tabular}
\end{table}


\subsection{Experimental Datasets}

\noindent\textbf{Wearable HAR dataset~(w-HAR)~\cite{w_har}:} The w-HAR dataset contains data of 22 users performing seven (\textit{jump, lie down, sit, stand, stairs down, stairs up, and walk}) and transitions between them. The data is collected using an 3-axis accelerometer and a wearable stretch sensor. The raw data from the users is segmented into activity windows and labeled such that each window contains a single activity. Overall, the dataset contains 4470 windows. Using these windows, the dataset generates discrete wavelet transform~(DWT) of the accelerometer data and fast Fourier transform~(FFT) of the stretch sensor data. These DWT and FFT features are then supplemented with the minimum and maximum values of the stretch sensor data, and activity window duration to generate 120 features for each window.
We use the default feature set provided by the dataset in our analysis.

\begin{figure*}[t]
    \centering
    \includegraphics[width=1\linewidth]{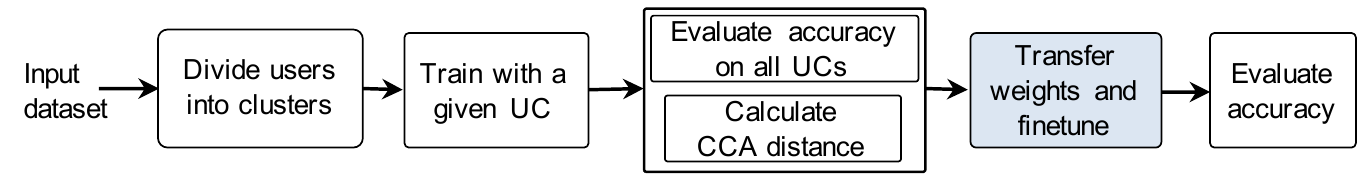}
    \caption{Overview of the transfer learning approach for HAR}
    \label{fig:tl_flow}
    \vspace{-3.5mm}
\end{figure*}

\noindent\textbf{UCI HAR dataset~\cite{anguita2013public}:} The UCI HAR consists of data from 30 users who performed six activities (\textit{lie down, sit, stand, stairs down, stairs up, and walk}) while wearing a smartphone on the waist. The dataset records readings from the accelerometer and gyroscope sensors in the smartphone. The dataset provides 561 time and frequency domain features for each activity window of 2.56~s. We use the default feature set provided by the dataset in our analysis.

\noindent\textbf{UCI HAPT dataset~\cite{reyes2016transition}:} This is an updated version of the UCI HAR dataset where the authors added information about postural transitions, such as stand-to-sit and sit-to-stand. Similar to the UCI HAR dataset, the HAPT dataset provides 561 time and frequency domain features for each activity.

\noindent\textbf{UniMiB dataset~\cite{micucci2017unimib}:} The UniMiB SHAR is a dataset of acceleration samples acquired with an Android smartphone. The dataset includes nine physical activities performed by 30 subjects. The dataset provides pre-processed windows for user activities. We generate DWT and FFT features, resulting in 450 features for each window.

\noindent\textbf{WISDM dataset~\cite{kwapisz2011activity}:} The WISDM dataset consists of data from 36 subjects who performed six activities (\textit{Walking, Jogging, Upstairs, Downstairs, Sitting}). The dataset records readings from the accelerometer and gyroscope sensors. Following the authors' recommendation, we use 10~s windows to generate 5424 activity windows from the raw data. Then, we produce the DWT and FFT features for the accelerometer data in each window, resulting in 405 features for each window.

\section{Transfer learning for HAR}
\subsection{Flow of the proposed transfer learning approach}

This subsection overviews the flow of the proposed approach illustrated in Figure~\ref{fig:tl_flow}. Then, the remaining subsections describe each step in more detail.

\begin{enumerate}
    \item The feature data in each dataset is split into multiple clusters using k-means clustering. The clustering ensures that users across different clusters are more dissimilar than the random partitioning.
    \item Train a CNN classifier for each user cluster (UC) obtained in step 1. The accuracy obtained in this step is the baseline accuracy for each UC, since the training data is available at design time.
    \item Evaluate the accuracy for unseen UCs using the classifiers trained in step 2. The accuracy obtained in this step is referred to as the \textit{cross-UC accuracy}. This step also calculates the CCA distance between trained CNNs.
    \item Transfer the CNN weights between UCs and fine-tune the layers that provide distinguishable information for each UC. Finally, evaluate the accuracy after fine-tuning the CNN layers.
\end{enumerate}

\noindent The rest of this section details these steps.

\subsection{Clustering users with distinct activity patterns}\label{sec:kmeans}

The first step in the proposed framework is partitioning the users into clusters. Clustering ensures that users in separate clusters have as distinct activity patterns as possible, as illustrated in Figure~\ref{fig:motivation}. Hence, we can analyze the benefits and efficiency of transfer learning under more challenging conditions than random partitioning. 
We employ the following steps to obtain the user clusters.

\begin{figure*}[t]
    \centering
    \includegraphics[width=1\linewidth]{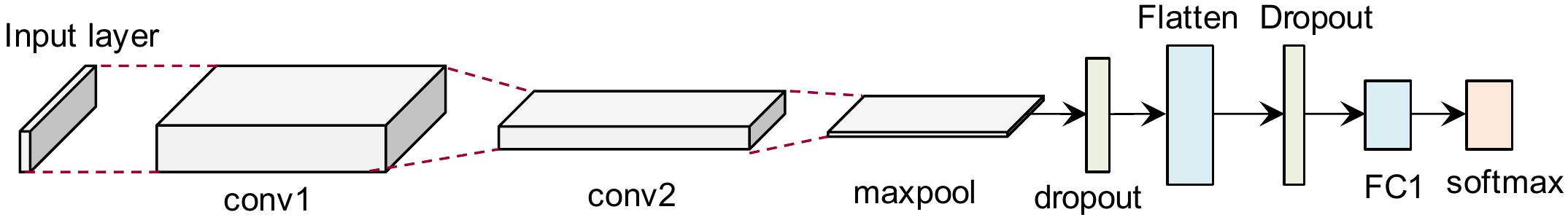}
    \caption{The architecture of CNN. The layers annotated at the bottom are used in CCA distance.}
    \label{fig:CNN_archi}
    \vspace{2mm}
    \includegraphics[width=1\linewidth]{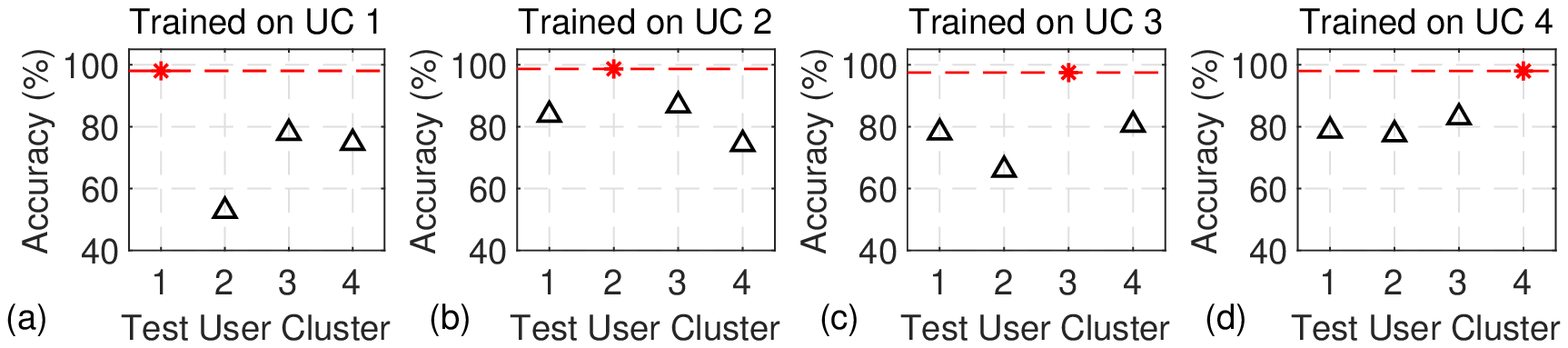}
    \caption{The accuracy of the CNNs tested with different UCs for the w-HAR dataset. The red star shows the accuracy of the UC used training while the triangles show cross-UC accuracy.}
    \label{fig:CNN_accuracy}
    
\end{figure*}

\noindent\textbf{Representative window for each user:} 
Each user has multiple windows for the same activity since the collected data is segmented before feature generation. For example, one-minute-long walking data is divided into 60 activity windows, assuming a one-second segment duration. Furthermore, activity data may be coming from different experiments with small differences in sensor locations. As a result, there are variations in the features across different windows, even for a given user and activity.  
To bypass the variations across windows and facilitate clustering, we identify a \textit{representative window} for each activity of each user. For example, consider our first dataset with 22 users and 8 activities. 
To obtain the representative windows we first extract all the activity windows for a given user and activity (e.g., all windows labeled as ``walk'' for user-1). Then, we compute the mean Euclidean distance from each window to all other windows for this user-activity pair. A large distance means that the corresponding window is more likely to be an outlier.
In contrast, the window with the smallest mean distance to the other windows is marked as the \textit{representative window} for the corresponding user-activity pair, as illustrated with red lines in Figure~\ref{fig:motivation}.

\noindent\textbf{k-means clustering:} 
The previous process results in one representative window for each activity for each user. That is, each of the 22 users have 8 representative windows (one for each activity) in our dataset with 22 users and 8 activities. 
Next, we compute the mean correlation distance~\cite{van2012metric}
between the representative windows across different users for each activity.
A small distance means that the activity pattern of the user is similar to other users.
Conversely, a larger distance implies that the user's activity pattern is separated from the other users. 
The distance to other users for each activity is stored as a multidimensional vector whose length is equal to the number of activities in the dataset. 
Finally, the distance vectors are used with the k-means algorithm~\cite{friedman2001elements} to generate user clusters that are as separated as possible.
Based on our empirical observations of the data, we choose four clusters each for the five datasets we use in this paper.

\subsection{Baseline classifier training for each user cluster}
\noindent\textbf{CNN for HAR:} We design a CNN for each input dataset to recognize the activities in the respective dataset, as shown in Figure~\ref{fig:CNN_archi}. For each dataset, the data dimensions are different while the structure remains the same. The input layer of the CNN takes the feature vectors as the input in the form of a 2-dimensional~(2D) image.
This is followed by two convolutional layers, max-pooling, and flatten layers. After flattening the data, we feed it to the two fully connected~(FC) layers before applying the softmax activation to classify the activity. Additional details on the structure of the CNN and training parameters are presented in Appendix A.


\noindent\textbf{CNN accuracy evaluation:} The CNN shown in Figure~\ref{fig:CNN_archi} is trained for each UC obtained in the previous section. We use 60\% data of each UC for training while 20\% data is used for cross-validation during training. We train 10 networks with each UC with different initialization such that we can analyze both CCA distance and accuracy on an average basis.
Next, we analyze the accuracy of the CNN for the UCs not seen during training using all the data of the unseen UCs. Figure~\ref{fig:CNN_accuracy} shows the average accuracy of CNNs trained with each of the four UCs obtained from our dataset and tested on the other three.
First, we see that the CNNs achieve a high accuracy on 20\% test data of the UC used for training. However, there is a significant reduction in cross-UC accuracy. For example, the accuracy for UC 2 when tested on CNNs trained with UC 1 is only about 52\%. 
Similar behavior is observed for other UCs as well, with the accuracy drop ranging from 10\%--40\%.
This shows that the CNN is only able to learn the data pattern of the current UC and it cannot generalize other UCs with distinct activity patterns.
In the next section, we analyze the distance between networks trained with different UCs to gain better insight into the representations learned by the CNN. Additional details on the accuracy and cross-UC accuracy of other datasets are presented in Appendix A.


\subsection{Analysis of distance between trained networks}

\noindent \textbf{Background on CCA Distance:} 
We employ CCA to analyze the distance between different networks and understand the representational similarity between network layers~\cite{morcos2018insights}.
It analyzes the representational similarity between networks by analyzing the ordered output activations of neurons on a set of inputs, instead of working on the network weights directly. 
Taking the activation vectors of neurons from two layers (trained with different UCs or with different initializations) as inputs, CCA first finds the linear combinations of the activations such that they are as correlated as possible.
Once the correlations are obtained, they are used to compute the distance between the two activation vectors, i.e., between the two layers~\cite{morcos2018insights}.
CCA has been successfully used to analyze neural network similarities for medical imaging~\cite{raghu2019transfusion}, language models~\cite{saphra2019understanding}, and speech recognition~\cite{qin2020towards}.
We use the implementation proposed in~\cite{morcos2018insights} to analyze the distance between the CNNs trained for HAR.
\begin{figure*}[h]
    \centering
    \includegraphics[width=1\linewidth]{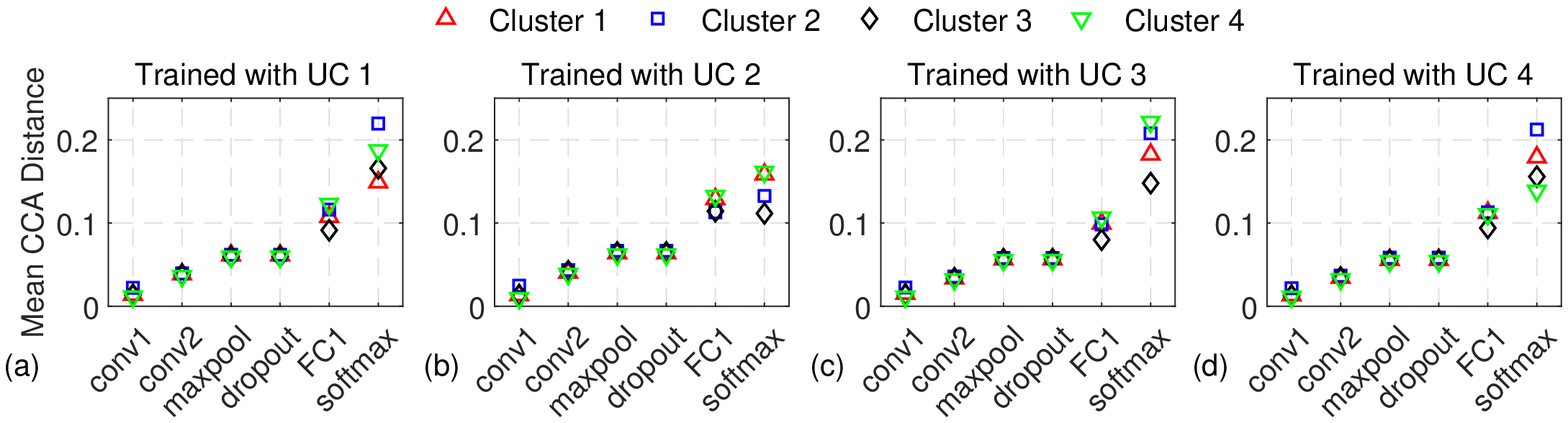}
    \caption{The CCA distance between CNNs trained with (a) UC 1, (b) UC 2, (c) UC 3, and (d) UC 4 from the w-HAR dataset when tested on all the four UCs.}
    \label{fig:cca within}
    \includegraphics[width=1\linewidth]{figures/accuracy_elab_ft.eps}
    \caption{Comparison of accuracy between original and fine-tuned CNN for the w-HAR dataset.}
    \label{fig:finetune accuracy}
    \vspace{-1mm}
\end{figure*}
\noindent\textbf{Distance between networks trained with same UC:} To analyze which layers generalize between users, we calculate the mean CCA distance between the networks trained with the same UC. To this end, 10 CNNs with different initializations are trained with the training data of each UC. Next, we find the mean pairwise CCA distance between the corresponding layers of the 10 CNNs trained with each UC.
Figure~\ref{fig:cca within} shows the mean CCA distance among NNs trained with 4 UCs with the w-HAR dataset.
Each sub-figure in Figure~\ref{fig:cca within} shows the distance between networks trained with a single UC when tested on all the four UCs. 
For example, Figure~\ref{fig:cca within}(a) shows the mean distance for the 10 networks when they are trained on UC 1 and tested on all four user clusters.
The figure shows that for the first four layers of the CNN~(two convolution layers, one max-pooling layer, and one dropout layer), the mean CCA distance is low regardless of the input UC.
Moreover, the distances are almost equal for all the four UCs. This shows that in the first four layers learn features that are similar for all the users in the dataset. To further validate this behavior, we calculate the pairwise CCA distance with features of (UC 1 + UC 2), (UC 1 + UC 3), and (UC 1 + UC 4) on networks trained with UC1. For each of these cases, the CCA distance follows a pattern similar to Figure~\ref{fig:cca within}. 

The networks start to diverge for different UCs from the first fully connected layer. The largest divergence is seen at the softmax layer where the distance is lowest for the UC used for training. 
This means that the fully connected layers extract information that is specific to each UC and do not generalize to other UCs.
We observe a similar trend for the other datasets as well.
In addition to this, we also calculate the CCA distance between networks trained with different UCs (\textit{e.g.,} the distance between CNNs trained with UC~1 and UC~2). The analysis of the distance from this perspective helps in understanding the similarity between networks trained with users that have distinct activity patterns. The details of this analysis are presented in Appendix B.
\subsection{Transferring the NN and fine-tuning}
CCA distance analysis reveals that the convolutional layers provide general features while the deeper layers provide the most distinguishing information. 
We use this insight to optimize the transfer learning process and improve the training time for new UCs. 
Specifically, we transfer the weights of the first four layers from a trained CNN to a network targeting another UC.
Then, the deeper layers are fine-tuned with the data of the new UC.
The fine-tuning process uses 60\% of the new UC's data for training, 20\% data for cross-validation and the remaining 20\% for testing.
Following this process, we are able to avoid training the convolutional layers, thus saving a significant amount of computations.
We evaluate the accuracy after fine-tuning under two scenarios: 1) fine-tune the last FC layer and 2)~fine-tune the last two FC layers.
Figure~\ref{fig:finetune accuracy} shows that the accuracy for new UCs improves significantly after fine-tuning either the last FC layer or last two FC layers for w-HAR dataset. With fine-tuning of one layer, we obtain on average 18\% accuracy improvement. 
Specifically, the accuracy for UC 2 improves from 52\%, 64\%, and 75\% to 90\%, 92\%, and 91\%, respectively. When we fine-tune the last two layers, the accuracy improves to 95\% for all UCs.




\section{Evaluations}
\label{sec:experimental_results}

\subsection{CNN training details}\label{sec:appendix_cnn}

We design a convolutional neural network~(CNN) for each of the input datasets to recognize the activities, as shown in Figure~4.
The feature vector is first converted to an image to be used as the input to the CNN. The dimensions of the image for each dataset are shown in the second column of Table~\ref{tab:CNNconfig}. 
The input layer is followed by two convolutional layers with 32 and 64 channels, respectively. After performing the convolutions, the data is passed to a max-pooling layer to extract the most distinguishable features.
It is then followed by a flattening layer that generates the input vector for the fully connected~(FC) layers. One FC layer with 128 neurons and relu activation is included in the CNN. 
To avoid excessive dependency on certain neurons, two dropout layers with a probability of 0.25 and 0.5 are employed after the max-pooling layer and fully connected layers, respectively.
Finally, an output layer with the softmax activation performs the activity classification. The dimensions of each CNN layer for the five datasets are shown in Table~\ref{tab:CNNconfig}.



We use Tensorflow~2.2.0~\cite{tensorflow2015-whitepaper} with Keras~~2.3.4~\cite{chollet2015keras} to train the CNNs. Categorical cross-entropy is employed as the loss during training. The algorithm used for training is the Adadelta~\cite{zeiler2012adadelta} optimizer with an initial learning rate of 0.001 and a decay rate of 0.95. We train the CNNs for 100 epochs using a batch size of 128. The training is performed on Nvidia Tesla GPU V100-SXM2 with 32~GB of memory. The details on training time are provided in Section~\ref{sec:training_time}.

\begin{table*}[h!]
\centering
\caption{Data dimensions for five datasets}
\label{tab:CNNconfig}
\setlength{\tabcolsep}{3pt}
\begin{tabular}{lcccccccc}
\toprule
         & Input   & Conv1    & Conv2    & Max-pooling & Flatten & FC1 & FC2 & Softmax \\ \midrule
w-HAR    & (4, 30, 1)  & (4, 30, 32)  & (2, 28, 64)  & (1, 14, 64)    & 896     & 128 & 8   & 8       \\
UCI HAR  & (33, 17, 1) & (33, 17, 32) & (31, 15, 64) & (15, 7, 64)    & 6720    & 128 & 6   & 6       \\
UCI HAPT & (33, 17, 1) & (33, 17, 32) & (31, 15, 64) & (15, 7, 64)    & 6720    & 128 & 12   & 12       \\
UniMiB & (25, 18, 1) & (25, 18, 32) & (23, 16, 64) & (11, 8, 64)    & 5632    & 128 & 9   & 9       \\
WISDM & (27, 15, 1) & (27, 15, 32) & (25, 13, 64) & (12, 6, 64)    & 4608    & 128 & 6   & 6       \\
\bottomrule
\end{tabular}
\end{table*}

\begin{figure*}[h]
    \centering
    \includegraphics[width=1\linewidth]{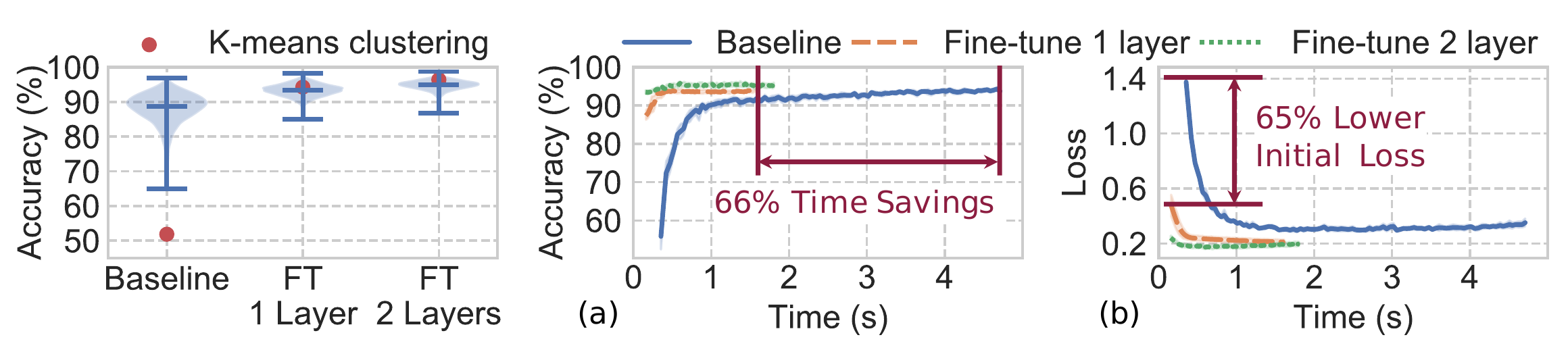}
    
    \begin{minipage}[h]{0.33\linewidth}
    \caption{Comparison of cross-UC accuracy between original and fine-tuned NN for 200 UCs from the \textit{w-HAR} dataset.}
    \label{fig:random_cluster}
    \end{minipage}
    \hspace{2mm}
    \begin{minipage}[h]{0.60\linewidth}
    \vspace{-4mm}
    \caption{Transfer learning improvement analysis: (a) Training time, (b) Loss for the \textit{w-HAR} dataset}
    \label{fig:Loss}
    \end{minipage}

    \includegraphics[width=1\linewidth]{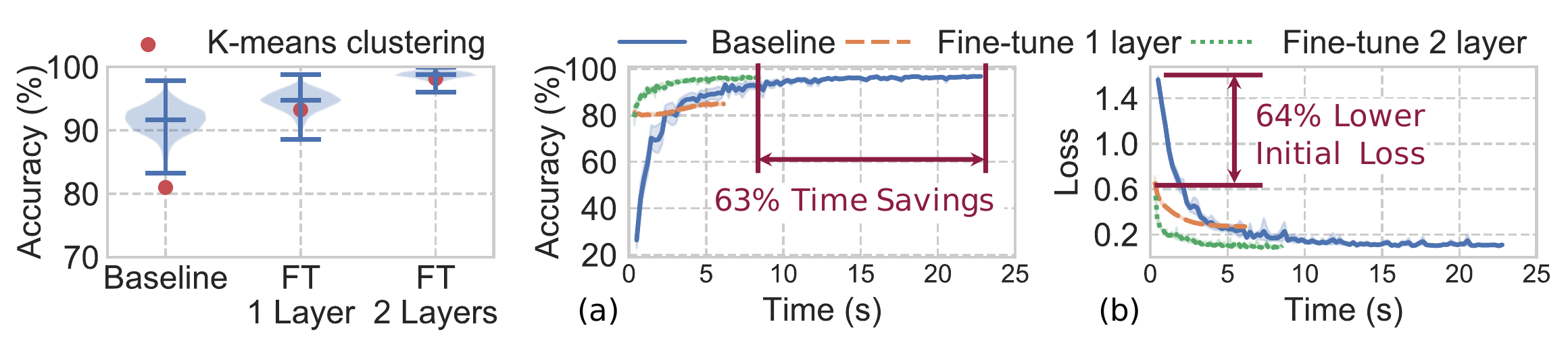}
    
    \begin{minipage}[h]{0.31\linewidth}
    \caption{Comparison between original and fine-tuned NN for 100 UCs from the \textit{UCI HAR} dataset.}
    \label{fig:random_cluster_uci}
    \end{minipage}
    \hspace{3mm}
    \begin{minipage}[h]{0.6\linewidth}
    \vspace{-4mm}
    \caption{Transfer learning improvement analysis:(a) Training time, (b) Loss for the \textit{UCI HAR} dataset.}
    \label{fig:Loss_UCI}
    \end{minipage}

    \includegraphics[width=1\linewidth]{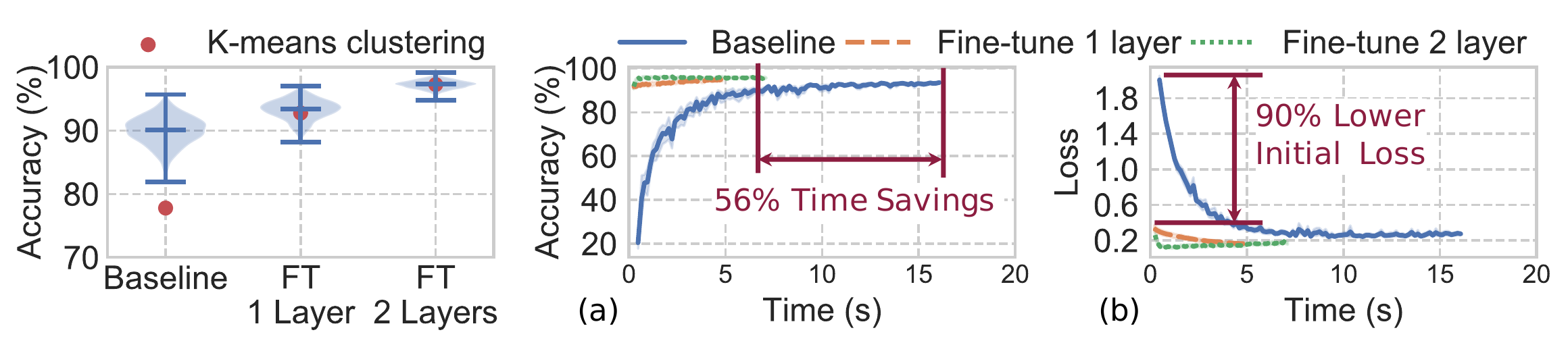}
    
    \begin{minipage}[h]{0.31\linewidth}
    \caption{Comparison between original and fine-tuned NN for 100 UCs from the \textit{UCI HAPT} dataset.}
    \label{fig:random_cluster_hapt}
    \end{minipage}
    \hspace{3mm}
    \begin{minipage}[h]{0.6\linewidth}
    \vspace{-4mm}
    \caption{Transfer learning improvement analysis:(a) Training time, (b) Loss for the \textit{UCI HAPT} dataset.}
    \label{fig:Loss_HAPT}
    \end{minipage}
\end{figure*}

\subsection{Accuracy analysis}
The analysis performed in the previous sections showed results with user clusters that are designed to be as distinct as possible. In this section, we validate the transfer learning approach on randomly generated user clusters. 
To this end, we first generate 200 random user splits from the w-HAR dataset. Each user split contains four user clusters, in line with previous sections.
We then train 5 CNNs for each cluster and test their cross-UC accuracy before applying transfer learning.
The first column in Figure~\ref{fig:random_cluster} shows the distribution of the cross-UC accuracy. 
The minimum cross-UC accuracy is 65\% for all the UCs among 200 random user splits. We also show the minimum accuracy for the k-means clustering using a red dot. The minimum accuracy with k-means is 52\%, which is lower than the minimum accuracy for all random user splits. This experimentally shows that our k-means clustering approximates the \textit{worst-case scenario} well, where the users are as distinct as possible.
Next, we fine-tune the last one and two layers of the CNNs to capture the information specific to each user cluster, as shown in the second and third columns of Figure~\ref{fig:random_cluster}, respectively.
We also analyze the accuracy for 100 randomly chosen UCs and convergence for the UCI HAR and UCI HAPT datasets, respectively. Figures~\ref{fig:random_cluster_uci} and~\ref{fig:random_cluster_hapt} show that the median accuracy obtained for the 100 random clusters is similar to the accuracy obtained from the k-means clustering. This is in line with the results from the w-HAR dataset.
The classification accuracy improves significantly after the fine-tuning process. Specifically, the median accuracy after fine-tuning one layer is 93\%, while it further increases to about 95\% by fine-tuning last two layers. These median accuracies are very close to the accuracy obtained for the k-means clustering. This shows our k-means clustering is representative of a wide range of randomly generated user clusters.
We also note that some UCs achieve a higher accuracy after fine-tuning when compared to the k-means clustering. This is because the users in these clusters have similar features. Conversely, some UCs have lower accuracy after fine-tuning because the source UC does not have all the activities present in the target UC.
We also provide additional details on the CNN accuracy for UCI HAR and UCI HAPT datasets, respectively.
Figure~\ref{fig:CNN_accuracy_UCI} shows the average accuracy of CNNs trained with each of the four UCs obtained from UCI HAR and tested on the other three.
The CNNs achieve high accuracy on the 20\% test data of the UC used for training. However, there is a reduction in the cross-UC accuracy. This is in line with the behavior on the w-HAR dataset. 
Next, Figure~\ref{fig:finetune_accuracy_uci} shows the accuracy after fine-tuning the CNNs with the data of unseen UCs for the UCI HAR dataset.
As expected, the recognition accuracy improves when we fine-tune either the last FC layer or last two FC layers. On average, the accuracy improves by 10\% when we fine-tune the last two FC layers.
Similarly, Figures~\ref{fig:CNN_accuracy_HAPT} and~\ref{fig:finetune_accuracy_hapt} show the accuracy for the UCI HAPT dataset before and after fine-tuning, respectively. We observe results similar to the UCI HAR dataset. The average accuracy improvement after fine-tuning the last two FC layers is 10\%. 
Figures~\ref{fig:CNN_accuracy_UniMiB} and~\ref{fig:finetune_accuracy_UniMiB} show the accuracy for the UniMiB before and after fine-tuning, respectively. We observe that cross-UC accuracy before fine-tuning is low. The lowest cross-UC accuracy is about 7\%. This is because the user data pattern is very different than other datasets. After 2 layers of fine-tuning, all cross-UC accuracies increase up to 88\%. 
Figures~\ref{fig:CNN_accuracy_WISDM} and~\ref{fig:finetune_accuracy_WISDM} show the accuracy for the WISDM before and after fine-tuning, respectively. We observe results similar to the UCI HAR dataset.
In summary, the proposed transfer learning approach of fine-tuning the deeper layers of the network significantly improves the classification accuracy. This shows that the proposed transfer learning approach provides accuracy improvements for all the five datasets used in our work.


\subsection{Training time, loss and convergence analysis}
\label{sec:training_time}
Transfer learning provides benefits in training speed and convergence when compared to training from scratch for new user clusters. Figure~\ref{fig:Loss}(a) shows the comparison of training time between the baseline and proposed transfer learning approach.
The transfer learning approach has both a higher starting accuracy and lower convergence time with respect to the baseline. Specifically, fine-tuning one layer converges to 93\% in about 1.6~s, which is 66\% lower than the baseline approach of training from scratch. When two layers of the CNN are fine-tuned, the accuracy is higher than the baseline with a small increase in the training time. 
Similar results are observed for the loss in Figure~\ref{fig:Loss}(b) where the starting loss with transfer learning is 65\% lower when compared to the baseline. The loss at the end of training is also lower with the transfer learning approach.
Figure~\ref{fig:Loss_UCI}(a) shows the comparison of training time between the baseline and proposed transfer learning approach.
The transfer learning approach has both a higher starting accuracy and lower convergence time compared to the baseline. Specifically, when two layers of the CNN are fine-tuned, the accuracy is the same as the baseline with 63\% lower training time. 
Similar results are observed for the loss in Figure~\ref{fig:Loss_UCI}(b) where the starting loss with transfer learning is 64\% lower than the baseline. The loss at the end of training is also lower with the transfer learning approach.
Figure~\ref{fig:Loss_HAPT}(a) and Figure~\ref{fig:Loss_HAPT}(b) show the corresponding results for the UCI HAPT data set. In this case, we observe that the training time is 56\% lower, while the starting loss is 90\% lower. 
In summary, this shows that the general features transferred from a trained CNN aid in learning the activities of new users.

\subsection{Energy and power consumption analysis}
This section analyzes the energy and power consumption of the proposed approach since the ability to run on low-power mobile devices is critical for its practicality. To enable this analysis, we implemented it on the Nvidia Jetson Xavier NX Development Kit~\cite{Xavier}. This board has a 6-core ARM CPU, 384 Nvidia CUDA cores, and 48 tensor processing units.
\begin{table*}[h]
\centering
\caption{Summary of power and energy consumption of the proposed approach when compared to training from scratch. Training from scratch is used as the baseline for power and energy saving calculations.}
\renewcommand\arraystretch{1}
\setlength{\tabcolsep}{4.5pt}
\footnotesize
\label{tab:power}
\begin{tabular}{lrrrrrrrrr}
\toprule
     & \multicolumn{3}{c}{Training from scratch}
     & \multicolumn{3}{c}{Fine-tune 1 layer}
     & \multicolumn{3}{c}{Fine-tune 2 layers}  \\
     \cmidrule(lr){2-4}
     \cmidrule(lr){5-7}
     \cmidrule(lr){8-10}
     & \multicolumn{1}{c}{\begin{tabular}[c]{@{}c@{}}w-HAR\end{tabular}} & \multicolumn{1}{c}{\begin{tabular}[c]{@{}c@{}}UCI-HAR\end{tabular}} & \multicolumn{1}{c}{\begin{tabular}[c]{@{}c@{}}WISDM\end{tabular}} & \multicolumn{1}{c}{\begin{tabular}[c]{@{}c@{}}w-HAR\end{tabular}} & \multicolumn{1}{c}{\begin{tabular}[c]{@{}c@{}}UCI-HAR\end{tabular}} & \multicolumn{1}{c}{\begin{tabular}[c]{@{}c@{}}WISDM\end{tabular}}& \multicolumn{1}{c}{\begin{tabular}[c]{@{}c@{}}w-HAR\end{tabular}} & \multicolumn{1}{c}{\begin{tabular}[c]{@{}c@{}}UCI-HAR\end{tabular}} & \multicolumn{1}{c}{\begin{tabular}[c]{@{}c@{}}WISDM\end{tabular}}\\
     \cmidrule(lr){2-4}
     \cmidrule(lr){5-7}
     \cmidrule(lr){8-10}
     \textbf{Configurations} & \multicolumn{3}{c}{4 UCs} & \multicolumn{3}{c}{4 UCs * 4 FT} & \multicolumn{3}{c}{4 UCs * 4 FT}\\
     \cmidrule(lr){2-4}
     \cmidrule(lr){5-7}
     \cmidrule(lr){8-10}
\textbf{CPU+GPU Power (W)}        & 1.93         &   5.20    &  4.69         & 1.27       &   2.40    &  2.26       & 1.30       &   3.81       &  2.29       \\ \midrule
\textbf{Total SoC Power (W)}      & 3.26         &   7.15    &  6.56         & 2.51       &   4.10    &   3.91      & 2.58       &   5.55       &   3.98      \\ \midrule
\textbf{Exec. Time per Case~(s)} & 28.08       &   59.66    &   28.80        & 17.72     &  33.48     &   18.50      & 18.74     &    37.84      &   20.12      \\ \midrule
\textbf{Energy per Case (J)}      & 91.76       &  427.00     &  189.15         & 44.63      &  137.33     &  72.42       & 48.45         &   210.01       &  80.28       \\ \midrule
\textbf{Power Saving (\%)}        &             & N/A       &           & 23    &  43     &   40      & 21    &  22        &  39       \\ \midrule
\textbf{Energy Saving (\%)}       &             & N/A       &          & 51     &  68     &   62     & 47     &  51        &   58     \\ \bottomrule
\end{tabular}
\end{table*}

\begin{table*}[h]
\centering
\caption{Summary of power and energy consumption of the proposed approach when compared to training from scratch. Training from scratch is used as the baseline for power and energy saving calculations.}
\renewcommand\arraystretch{1}
\setlength{\tabcolsep}{4.5pt}
\footnotesize
\label{tab:power_appendix}
\begin{tabular}{lrrrrrr}
\toprule
     & \multicolumn{2}{c}{Training from scratch}
     & \multicolumn{2}{c}{Fine-tune 1 layer}
     & \multicolumn{2}{c}{Fine-tune 2 layers}  \\
     \cmidrule(lr){2-3}
     \cmidrule(lr){4-5}
     \cmidrule(lr){6-7}
     &  \multicolumn{1}{c}{\begin{tabular}[c]{@{}c@{}}UCI-HAPT\end{tabular}} & \multicolumn{1}{c}{\begin{tabular}[c]{@{}c@{}}UniMiB\end{tabular}} & \multicolumn{1}{c}{\begin{tabular}[c]{@{}c@{}}UCI-HAPT\end{tabular}} & \multicolumn{1}{c}{\begin{tabular}[c]{@{}c@{}}UniMiB\end{tabular}}& \multicolumn{1}{c}{\begin{tabular}[c]{@{}c@{}}UCI-HAPT\end{tabular}} & \multicolumn{1}{c}{\begin{tabular}[c]{@{}c@{}}UniMiB\end{tabular}}\\
     \cmidrule(lr){2-3}
     \cmidrule(lr){4-5}
     \cmidrule(lr){6-7}
     \textbf{Configurations} & \multicolumn{2}{c}{4 UCs} & \multicolumn{2}{c}{4 UCs * 4 FT} & \multicolumn{2}{c}{4 UCs * 4 FT}\\
     \cmidrule(lr){2-3}
     \cmidrule(lr){4-5}
     \cmidrule(lr){6-7}
\textbf{CPU+GPU Power (W)}        & 5.47         &   4.41    &  2.52         & 2.00       &   3.21    &  2.16\\ \midrule
\textbf{Total SoC Power (W)}      & 7.36         &   6.34    &  4.24         & 3.54       &   4.92    &   3.86\\ \midrule
\textbf{Exec. Time per Case~(s)} & 58.67       &   39.97    &   32.79        & 22.39     &  37.00     &   24.72\\ \midrule
\textbf{Energy per Case (J)}      & 431.89       &  253.38     &  139.04         & 79.39      &  182.05     &  95.44\\ \midrule
\textbf{Power Saving (\%)}        &             & N/A       &    42       & 33    &  44     &   39\\ \midrule
\textbf{Energy Saving (\%)}       &             & N/A       &    68      & 58     &  69     &   62\\ \bottomrule
\end{tabular}
\end{table*}

\noindent\textbf{Methodology:} We evaluate the power and energy consumption of training the CNN under the following scenarios:
\begin{enumerate}
    \item Train the classifier on the board from scratch for all 4 UCs in each dataset. Then, find the average energy and power consumption.
    
    \item Fine-tune only one or two layers using the proposed transfer learning approach. Repeat the experiments for all UC combinations and datasets.
    
\end{enumerate}

In both cases, we repeat the training ten times and take the average to suppress runtime variations.

\noindent\textbf{Results:}
Table~\ref{tab:power} shows the power and energy consumption comparisons for w-HAR, UCI, and WISDM datasets, while the results for other datasets are in Appendix C. The proposed transfer learning approach achieves 21\%--43\% and 47\%--68\% reduction in the power and energy consumption, respectively. For instance, if we fine-tune one layer for the UCI-HAR dataset instead of training from scratch, we achieve 68\% savings in energy consumption.
Table~\ref{tab:power_appendix} shows the power and energy consumption comparisons for UCI-HAPT and UniMiB datasets. The proposed transfer learning approach achieves 33\%--44\% and 58\%--59\% reduction in the power and energy consumption, respectively.
These results show that the transfer learning approach shown in this paper provides an efficient mechanism to adapt HAR classifiers for new users among all datasets.
	\section{Conclusion}
HAR has wide-ranging applications in movement disorders, rehabilitation, and activity monitoring. 
This paper presented a transfer learning framework to adapt HAR classifiers to new users with distinct activity patterns.
We started with a representational analysis of CNNs trained with distinct user clusters. This analysis revealed that initial layers of the network generalize to a wide range of users while the deeper layers providing distinguishing information.
Based on this insight, we fine-tuned deeper layers of the network while transferring the other layers. 
Experiments on five HAR datasets showed that our approach achieves up to 43\% and on average 14\% accuracy improvement when compared to the accuracy without using transfer learning. 
Fine-tuning the deeper layers also leads to 66\% savings in training time, 68\% savings in energy and better convergence while maintaining the same accuracy as training from scratch for new users.

	\bibliographystyle{ACM-Reference-Format}
	\bibliography{references/embedded_refs,references/flexible,references/wearable_iot,references/health_refs,references/transfer_learning}
	
	\clearpage
	
\appendix
\renewcommand{\thesection}{Appendix \Alph{section}}
\section{Additional CCA distance analysis}\label{sec:cca_appendix}
\begin{figure*}[h]
    \centering
    \includegraphics[width=1\linewidth]{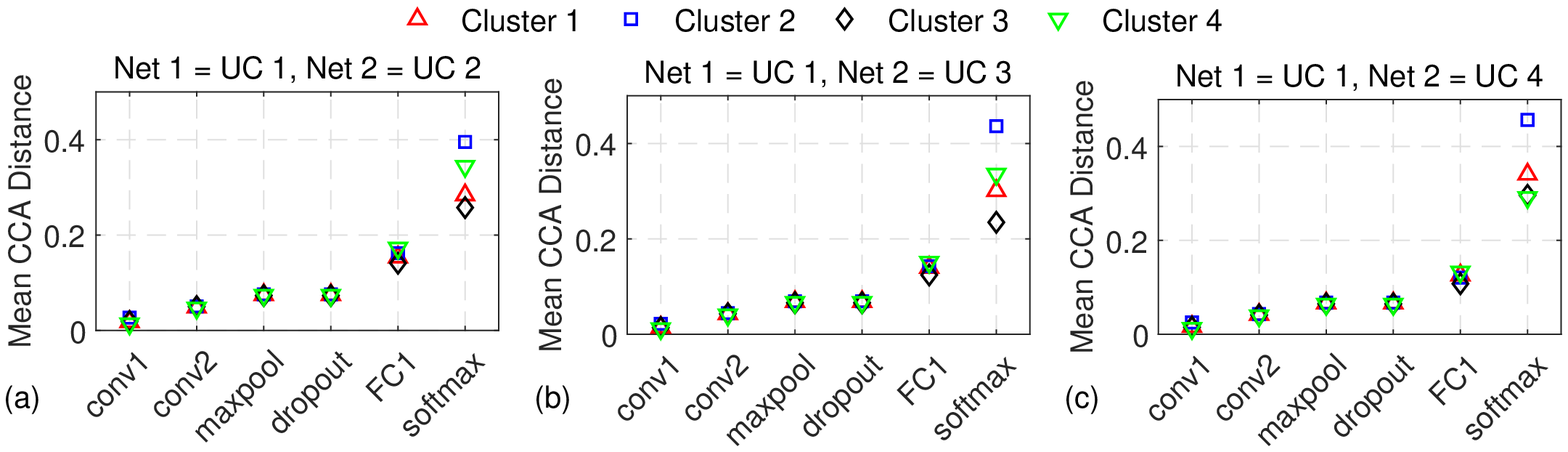}
    \caption{The CCA distance between CNNs trained with (a) UC 1 and UC 2, (b) UC 1 and UC 3, (c) UC 1 and UC 4 from the w-HAR dataset when tested on all the four UCs.}
    \label{fig:cca_cross_elab}
    \includegraphics[width=1\linewidth]{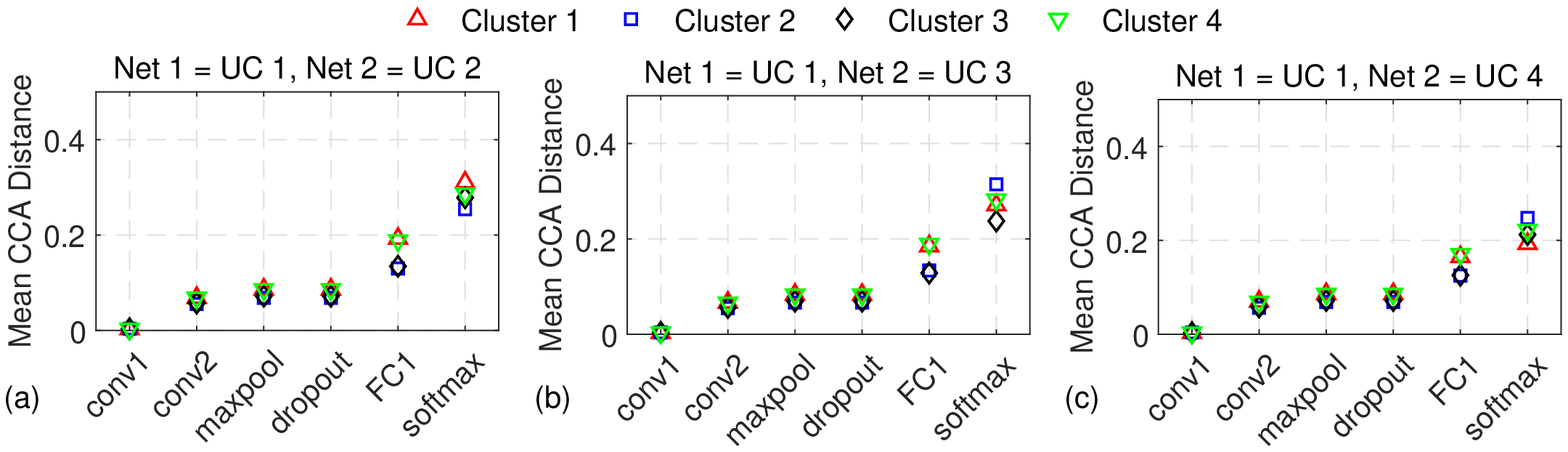}
    \caption{The CCA distance between CNNs trained with (a) UC 1 and UC 2, (b) UC 1 and UC 3, (c) UC 1 and UC 4 from the UCI HAR dataset when tested on all the four UCs.}
    \label{fig:cca_cross_uci}
    \includegraphics[width=1\linewidth]{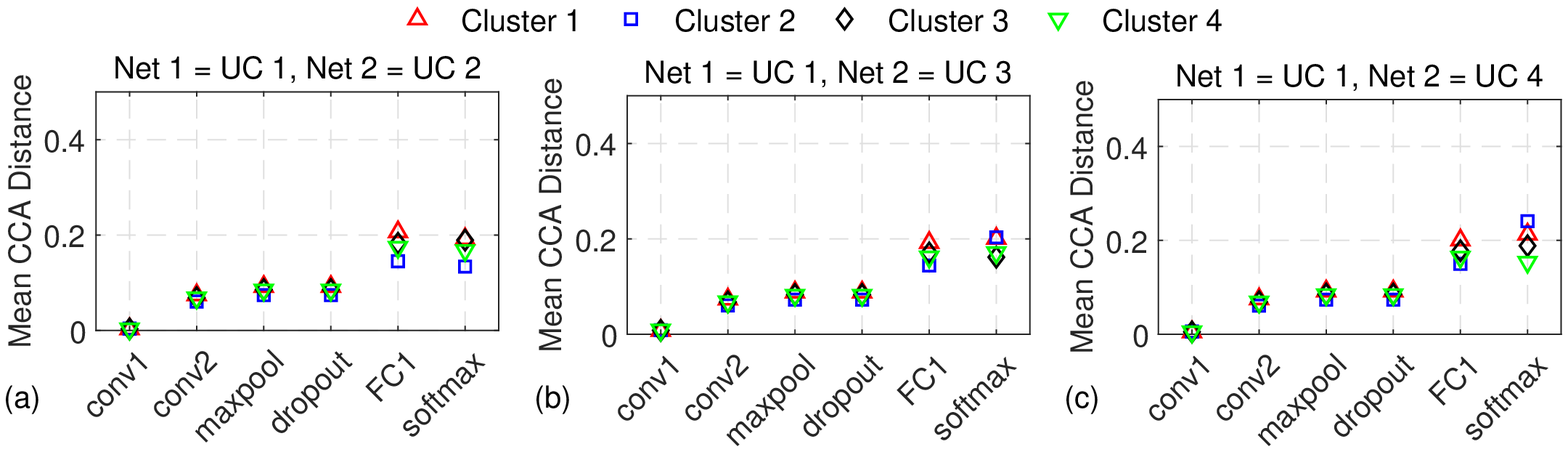}
    \caption{The CCA distance between CNNs trained with (a) UC 1 and UC 2, (b) UC 1 and UC 3, (c) UC 1 and UC 4 from the UCI HAPT dataset when tested on all the four UCs.}
    \label{fig:cca_cross_hapt}
\end{figure*}

\begin{figure*}[h]
    \includegraphics[width=1\linewidth]{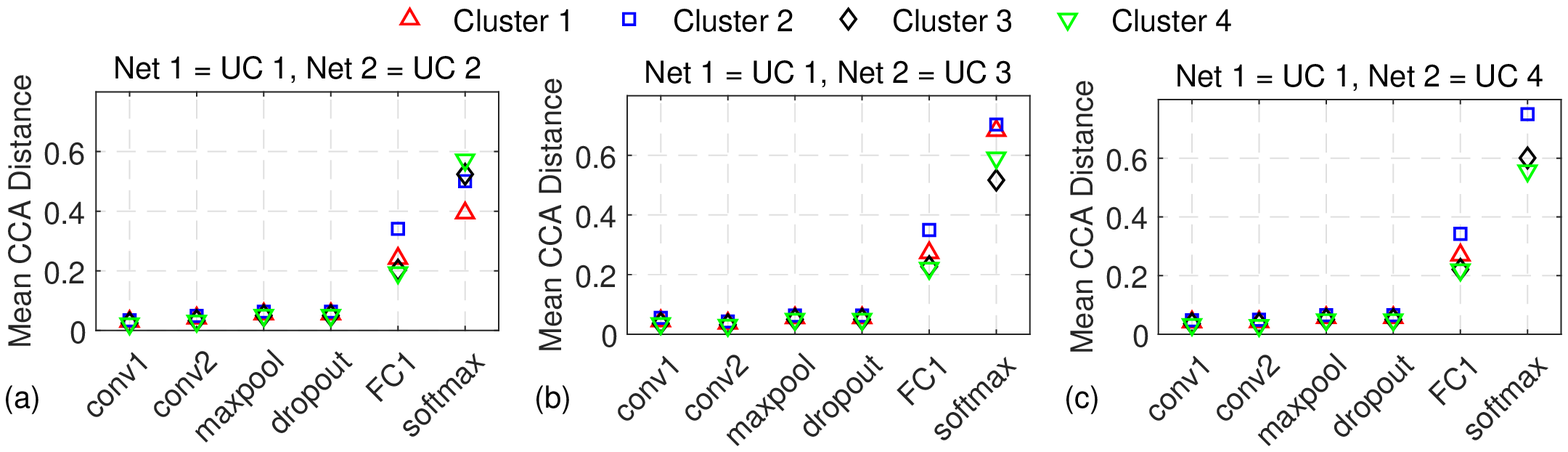}
    \caption{The CCA distance between CNNs trained with (a) UC 1 and UC 2, (b) UC 1 and UC 3, (c) UC 1 and UC 4 from the UniMiB dataset when tested on all the four UCs.}
    \label{fig:cca_cross_UniMiB}
    \includegraphics[width=1\linewidth]{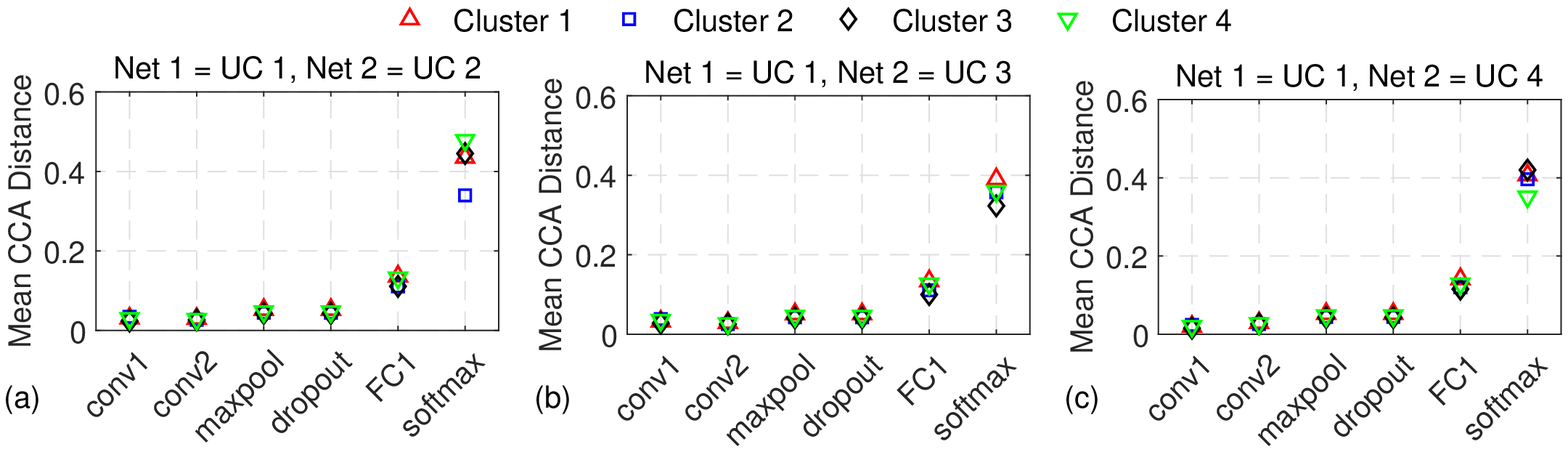}
    \caption{The CCA distance between CNNs trained with (a) UC 1 and UC 2, (b) UC 1 and UC 3, (c) UC 1 and UC 4 from the WISDM dataset when tested on all the four UCs.}
    \label{fig:cca_cross_WISDM}
\end{figure*}

This section describes the calculation of the CCA distance between networks trained with different UCs (\textit{e.g.,} the distance between CNNs trained with UC~1 and UC~2).
The analysis of the distance from this perspective helps in understanding the similarity between networks trained with users that have distinct activity patterns.
Figure~\ref{fig:cca_cross_elab} shows the CCA distance between CNNs trained with UC 1 and the other three UCs from the w-HAR dataset. We see that the first four layers are similar to each other regardless of the input data. Furthermore, the last layer shows the highest divergence in the CCA distances when tested with the four UCs. 
Figures~\ref{fig:cca_cross_uci}, ~\ref{fig:cca_cross_hapt}, ~\ref{fig:cca_cross_UniMiB}, and~\ref{fig:cca_cross_WISDM} show the CCA distances for CNNs trained with UCI HAR, UCI HAPT , UniMiB, and WISDM datasets, respectively. A similar pattern is observed for the UCI HAR, UCI HAPT , UniMiB, and WISDM datasets. The absolute distances are lower since there is a higher similarity between activity patterns of different UCs in the UCI datasets compared to the w-HAR dataset. For the UniMiB and WISDM datasets, the absolute distance is higher since there is lower similarity between the activity patterns of different UCs compared to the w-HAR dataset.

In summary, the additional results show that the first four layers learn general features that are applicable to all the UCs even when different UCs are used to train the CNN. 
Figures~\ref{fig:cca_cross_uci}, ~\ref{fig:cca_cross_hapt}, ~\ref{fig:cca_cross_UniMiB}, and~\ref{fig:cca_cross_WISDM} are provided starting with the next page.

\section{Source code}
The source code associated with the paper is available for download at \url{https://bit.ly/37sEehO}.

\begin{figure*}[h]
    \centering
    \includegraphics[width=0.95\linewidth]{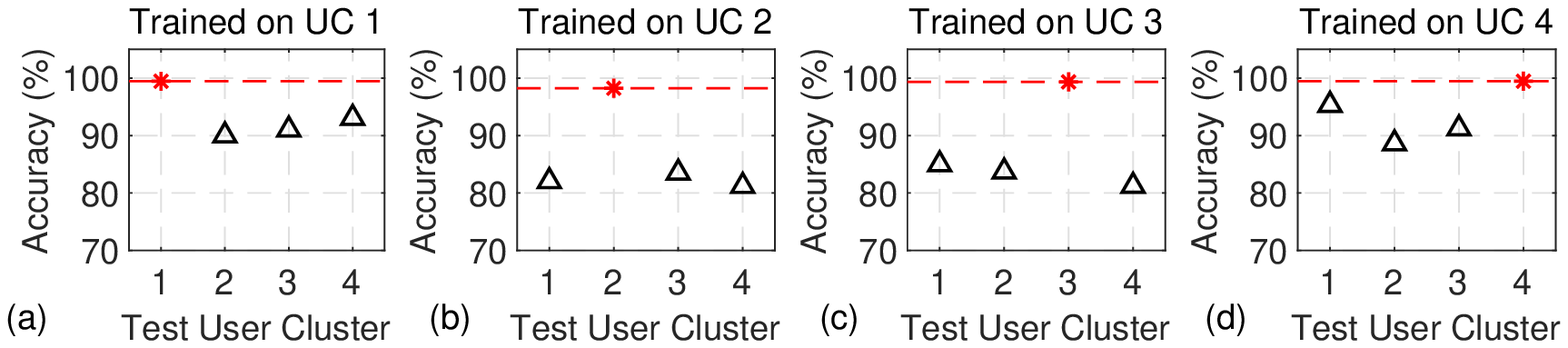}
    \caption{The accuracy of the CNNs tested with different UCs for the UCI HAR dataset. The red star shows the accuracy of the UC used training while the triangles show cross-UC accuracy.}
    \label{fig:CNN_accuracy_UCI}
    \includegraphics[width=1\linewidth]{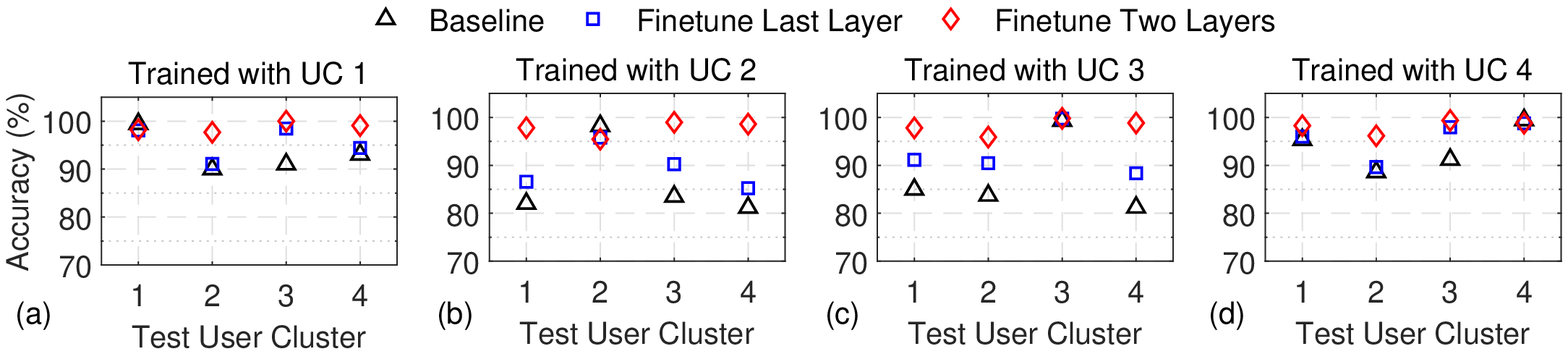}
    \caption{Comparison of accuracy between original and fine-tuned CNN for the UCI HAR dataset.}
    \label{fig:finetune_accuracy_uci}
    \includegraphics[width=0.95\linewidth]{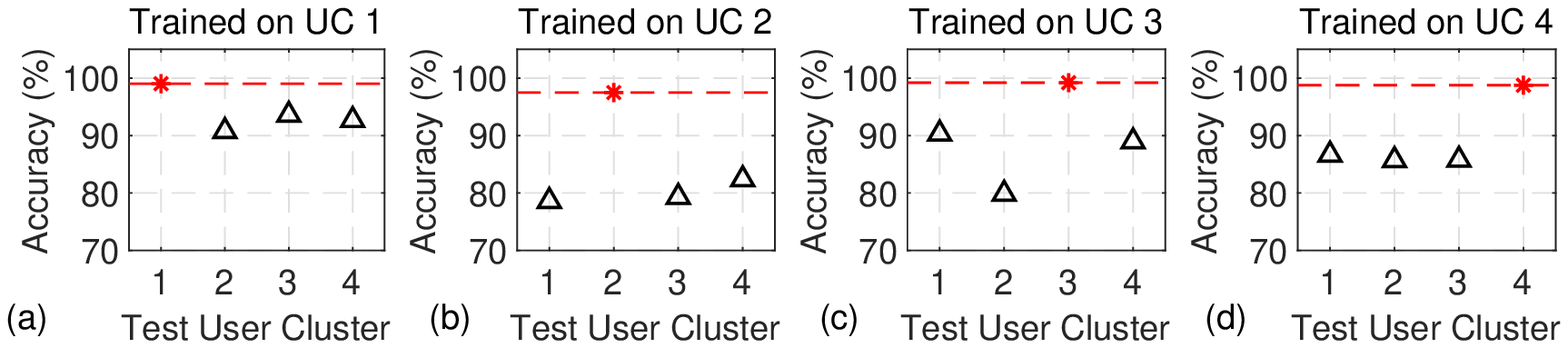}
    \caption{The accuracy of the CNNs tested with different UCs for the UCI HAPT dataset. The red star shows the accuracy of the UC used training while the triangles show cross-UC accuracy.}
    \label{fig:CNN_accuracy_HAPT}
    \includegraphics[width=1\linewidth]{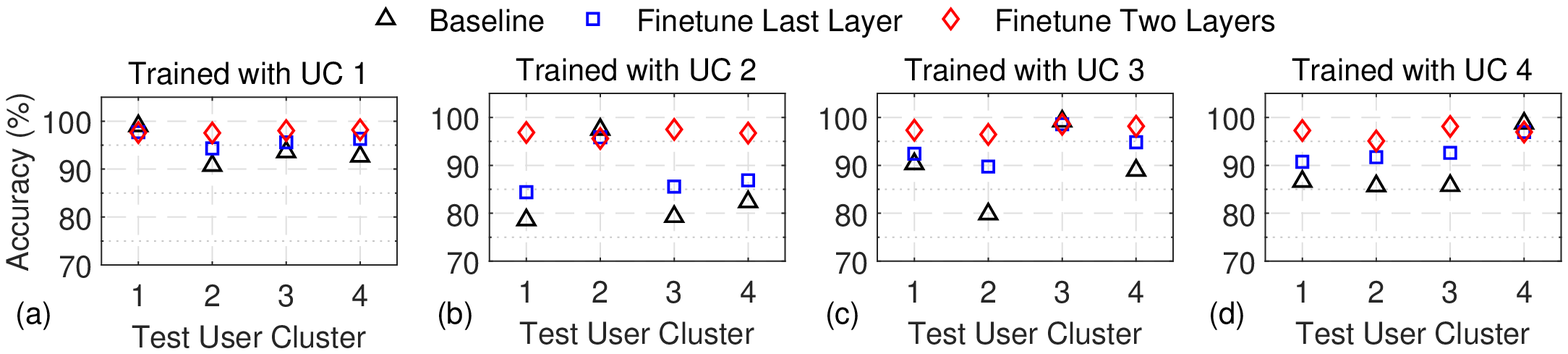}
    \caption{Comparison of accuracy between original and fine-tuned CNN for the UCI HAPT dataset.}
    \label{fig:finetune_accuracy_hapt}
\end{figure*}

\begin{figure*}[h]
    \includegraphics[width=0.95\linewidth]{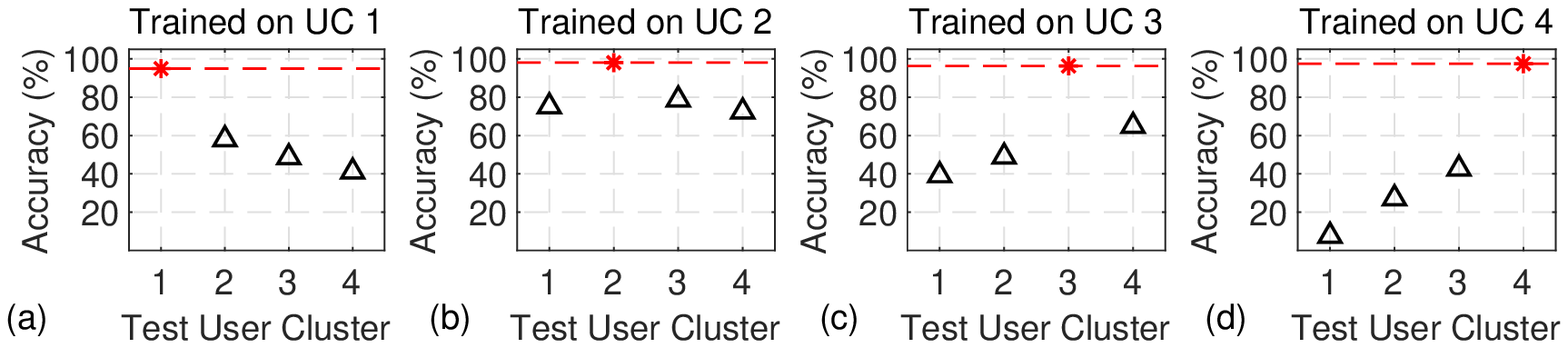}
    \caption{The accuracy of the CNNs tested with different UCs for the UniMiB dataset. The red star shows the accuracy of the UC used training while the triangles show cross-UC accuracy.}
    \label{fig:CNN_accuracy_UniMiB}
    \includegraphics[width=1\linewidth]{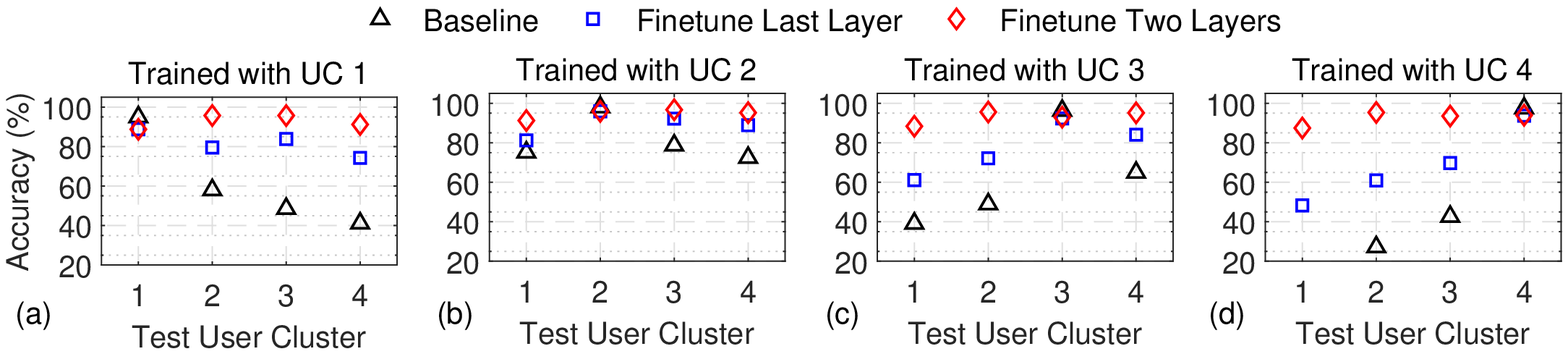}
    \caption{Comparison of accuracy between original and fine-tuned CNN for the UniMiB dataset.}
    \label{fig:finetune_accuracy_UniMiB}
    \includegraphics[width=0.95\linewidth]{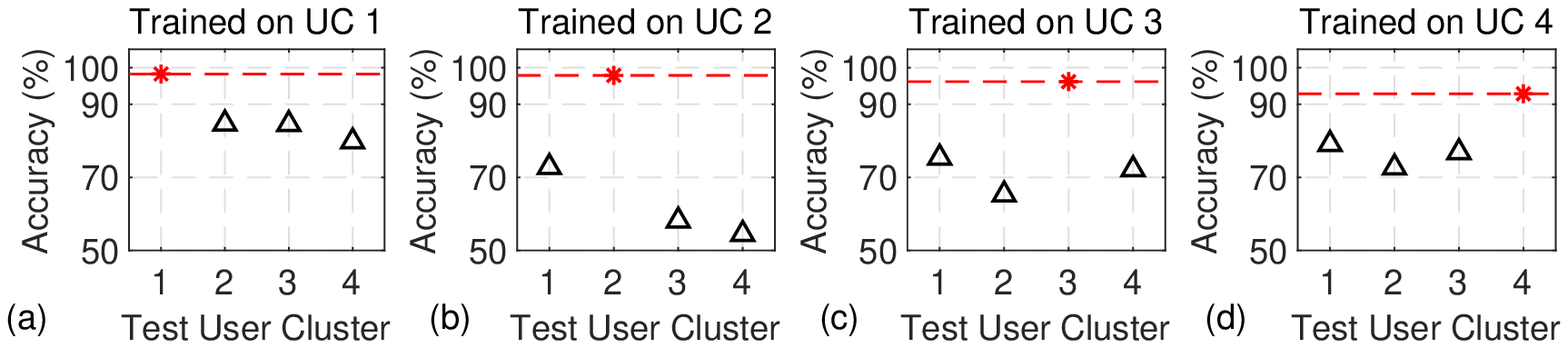}
    \caption{The accuracy of the CNNs tested with different UCs for the WISDM dataset. The red star shows the accuracy of the UC used training while the triangles show cross-UC accuracy.}
    \label{fig:CNN_accuracy_WISDM}
    \includegraphics[width=1\linewidth]{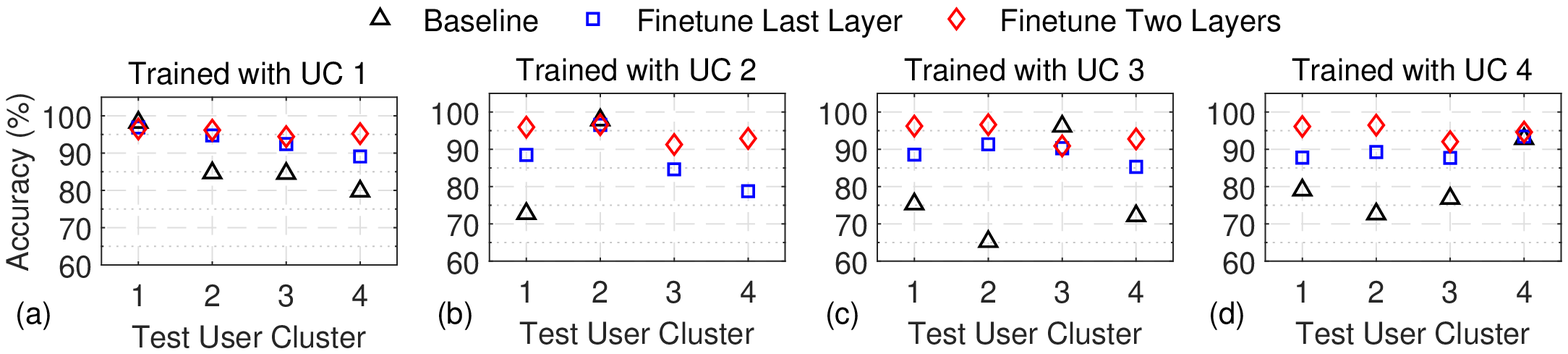}
    \caption{Comparison of accuracy between original and fine-tuned CNN for the WISDM dataset.}
    \label{fig:finetune_accuracy_WISDM}
\end{figure*}

\end{document}